\documentclass[twocolumn,nofootinbib,superscriptaddress,aps,prc]{revtex4-1}	
\usepackage{graphicx}
\usepackage{amsmath}
\usepackage[utf8]{inputenc}
\usepackage{latexsym}   
\usepackage{color}
\newcommand{\ave}[1]{\left\langle #1 \right\rangle}
\newcommand{\acf}[1]{\left\langle #1 \right\rangle_\mathrm{CF}}

\newcommand{\non}{\nonumber\\}

\begin{document}
\title{The effect of finite particle number sampling on baryon number fluctuations}
\author{Jan Steinheimer} \email{steinheimer@fias.uni-frankfurt.de}
\affiliation{Frankfurt Institute for Advanced Studies, Ruth-Moufang-Str. 1, 60438 Frankfurt am Main, Germany}
\author{Volker Koch} \email{vkoch@lbl.gov}
\affiliation{
Nuclear Science Division, Lawrence Berkeley National Laboratory,
Berkeley, California 94720, USA}

\begin{abstract}
The effects of finite particle number sampling on the net baryon number cumulants, extracted from fluid dynamical simulations, are studied. The commonly used finite particle number sampling procedure introduces an additional Poissonian (or multinomial if global baryon number conservation is enforced) contribution which increases the extracted moments of the baryon number distribution. If this procedure is applied to a fluctuating fluid dynamics framework one severely overestimates the actual cumulants. We show that the sampling of so called test-particles suppresses the additional contribution to the moments by at least one power of the number of test-particles. We demonstrate this method in a numerical fluid dynamics simulation that includes the effects of spinodal decomposition due to a first order phase transition.  Furthermore, in the limit where anti-baryons can be ignored, we derive analytic formulas which capture exactly the effect of particle sampling on the baryon number cumulants. These formulas may be used to test the various numerical particle sampling algorithms. 
\end{abstract}

\maketitle

The goal of heavy ion collision experiments is to study the properties of very hot and dense matter. It has been proposed that in the most energetic collisions of nuclei at the RHIC and LHC a new state of matter, the so called Quark-Gluon-Plasma, has been created \cite{Adams:2005dq,Back:2004je,Arsene:2004fa,Adcox:2004mh,Alt:2007aa}. Lattice QCD results predict the transition at vanishing net baryon number density to be a crossover \cite{Aoki:2006we,Aoki:2006br,Karsch:2001cy}. Due to the fermion sign problem, these calculations cannot be directly extended to the interesting region of high net baryon density where the cross-over may change to a first order phase transition. One of the main goals of current and future experimental programs is to find experimental signals for a possible first order phase transition and critical point in the phase diagram of the strong interaction, Quantum Chromo Dynamics (QCD). 

The systems created in these nuclear collisions are very small, rapidly expanding and not always in local thermal and chemical equilibrium. Thus the dynamical evolution of such a collision is far from trivial and sophisticated transport models are being employed. The current state-of-the-art of such models are the fluid dynamical hybrid models. In these models one uses a non-equilibrium initial state for a viscous fluid dynamical evolution, which is followed by a Boltzmann-transport description for the hadronic freeze-out phase. This setup is convenient as the fluid dynamical equations allow for a straight forward inclusion of the Equation of State (EoS).\\
Since the systems created in heavy ion collisions are rather small (on the order of $\sim$ 10-20 fm) the application of standard fluid dynamics is limited. In fluid dynamics one usually assumes that the particle number in a given fluid element sufficiently large so that local fluctuations of the particle number, and hence of the baryon- and energy density, can be neglected. Since this is not strictly the case in heavy ion collisions new fluid dynamical approaches, including local (thermal) fluctuations have been proposed \cite{Nahrgang:2011mg,Plumberg:2017tvu,Young:2014pka,Albright:2015uua}.\\
Understanding and propagating fluctuations in fluid dynamics is important since the fluctuations of conserved charges are considered a very sensitive probe  for the QCD phase transition and the associated critical point \cite{Stephanov:2008qz,Koch:2008ia}. This is motivated by the observation that in the vicinity of  the critical point  the correlation length diverges for an infinite system of matter. As a consequence also the higher order susceptibilities of e.g. the net baryon number will diverge at the critical end-point (and will show a strong increase in its vicinity). While the measured susceptibilities are very sensitive to the phase structure of QCD they are also affected by other aspects of the dynamical evolution, many of which have been discussed in recent literature \cite{Bzdak:2012ab,Bzdak:2016qdc,Kitazawa:2016awu,Feckova:2015qza,Begun:2004gs,Bzdak:2012an,Gorenstein:2008et,Gorenstein:2011vq,Sangaline:2015bma,Tarnowsky:2012vu,Xu:2014jsa,Adamczyk:2014fia,Adamczyk:2013dal}.\\
In order to achieve  a  realistic and quantitative description/prediction of these fluctuation observables it is important to include the critical phase transition dynamics in a dynamical model of the collision. Resent advances in that direction included an unstable equation of state in the standard fluid dynamics picture \cite{Steinheimer:2012gc}, allowing for a numerical 3+1D description of the spinodal decomposition at the first order phase transition. To include effects near the critical point it is necessary to also include thermal fluctuations in the standard fluid description. This has been done by amending the standard fluid dynamical equations by an equation of motion for the chiral field which serves as the order parameter of the phase transition \cite{Paech:2003fe,Nahrgang:2011mg,Herold:2013bi}. Other works also focus on the implementation of local thermal fluctuations in the dynamical fluid-evolution of the system.
 \cite{Plumberg:2017tvu,Young:2014pka,Albright:2015uua}.\\
A common denominator of all these fluid dynamical models is that at some point the transported fluid fields have to be converted into "real" hadrons. This is usually done via the Cooper-Frye (C-F) freeze-out prescription \cite{Cooper:1974mv}. In this paper we will discuss the effects of using the Cooper-Frye prescription on net baryon number and net proton number susceptibilities. In the first part we will introduce the most commonly used procedures. Then we will show the effect of finite particle sampling on the observed susceptibilities and will verify our analytically results with numerical simulations. Finally we will address and discuss the applicability of the Cooper-Frye picture and the fluctuating hydro framework on observable fluctuation measurements. 

\section{Particle production from fluid dynamics}
Fluid dynamical models are based on the conservation of energy momentum
\begin{equation}
\partial_{\mu} T^{\mu \nu} =0
\end{equation}
and the conservation of various charged currents
\begin{equation}\label{consr}
\partial_{\mu} j^{\mu}=0
\end{equation}
Here the net baryon number is the most commonly used conserved current, but in practice one could solve equation (\ref{consr}) for any number of conserved charges.
Note that in this picture the effects of finite number of particles in a given local cell $dV=dx^3$ are neglected, i.e. one essentially assumes that the particle number in a local cell is very large.
In reality the number of particles in a local cell is not very large, since the usual cell size is of the order of $1 \mathrm{fm}^{3}$ or less, $dx \leq 1$~fm. In order to accommodate effect of local thermal fluctuations, recently fluctuating hydro approaches have been developed. If done properly these models should correctly reproduce the thermal fluctuations of local cells. These models are especially useful as these thermal fluctuations will be the source of critical phenomena near the critical point and phase transition.\\
The result of a fluid dynamical simulation is the space-time  distributions of the energy-momentum-densities as well as the density of any conserved charges considered. To be able to compare such results with experimental measurements one has to translate these densities into distributions of real particles. This is typically achieved with the Cooper-Frye equation \cite{Cooper:1974mv}: 
\begin{equation}\label{cooper_frye}
E \frac{dN}{d^3p}=\int_\sigma f(x,p) p^\mu d\sigma_\mu \, .
\end{equation}
Here $f(x,p)$  corresponds Bose or Fermi single particle phase-space distribution functions. Typically one works in the grand canonical ensemble so that $f(x,p)$ depend essentially on the local temperature and chemical potentials, $T(x)$ and $\mu_{i}(x)$.\\
The Cooper-Frye equation \eqref{cooper_frye} relates the integral of the phase-space densities of particles passing through a freeze-out hyper-surface with the momentum distribution of these particles over the entire phase space. 
If there are no additional interactions past the freeze-out, all experimental cuts may then be performed on this momentum distribution. In practice, however, one has to consider additional features such as the decay of unstable resonances, re-scattering in a dilute hadronic phase and acceptance effects which affect particles differently depending on their position in phase space \cite{Werner:2010aa,Song:2013qma,Knospe:2015nva,Ryu:2012at,Steinheimer:2012rd}. Furthermore if particle fluctuations and correlations are to be investigated the final output of heavy ion simulation consists preferably of discrete particles with specific positions and momenta. To address these issues, one follows the C-F freeze-out with a transport evolution. In this case one needs the phase-space distribution of the hadrons right after freeze-out.  To achieve this,  
one usually selects a small sub-volume of the freeze-out hyper-surface -- typically one of the cells of the grid on which the fluid-dynamical variables are defined. Given such a cell one obtains the momentum spectra and the particle number from the underlying thermal distribution given the local temperature and chemical potentials and assigns the particles the spatial coordinate of the local cell. As already pointed out, the typical size of such a cell is usually so small that on the average it contains less than one particle. Nevertheless the commonly used particles samplings provide reliable information on event-averaged single particle distributions, which may then be compared with experiment. 
In addition algorithms have been developed to ensure the global conservation of conserved charges \cite{Petersen:2008dd}.\\ 
It is this type of ``freeze-out'' which provides the initial phase-space distribution for a transport evolution that we want to discuss further in this paper. In particular, we will demonstrate that the presently used sampling methods will lead to additional, potentially spurious contributions to particle number fluctuations and, consequently, to the extracted susceptibilities.

\section{The effects of finite particle number sampling}
Before we discuss the effect of finite particle sampling let us first clarify the concept of fluctuating fluid dynamics. Fluid dynamics, being a single-particle theory does not inherently carry information about fluctuations since it provides only information about mean values. However, one can consider an ensemble of fluid dynamics events which differ for example by their initial conditions. In this case the mean values for a given observable does fluctuate within this ensemble of events. One prominent example are the higher moments of the azimuthal distribution, $v_{n}$ \cite{Aguiar:2001ac,Kolb:2003zi}, which are well described by fluctuating initial conditions \cite{Schenke:2010rr,Werner:2010aa,Petersen:2010cw,Shen:2014vra} . Similarly, the inclusion of thermal fluctuations will generate an ensemble of fluid-dynamics events. Whatever the source of the fluctuations, at the end we will have an ensemble of fluid-dynamics events where, at freeze-out, the distribution of the various densities (energy, momentum and charge) will vary from event to event. Thus we will have an ensemble of freeze-out hyper-surfaces, where for examples the baryon number is distributed differently from event to event. Thus if we look at a subset of this hyper-surface, the baryon number will fluctuate although the total baryon number is conserved in the hydrodynamics evolution. In the following we will discuss how the various freeze-out schemes will map these fluctuation into the initial state of a transports evolution. 

As already discussed, typically the Cooper-Frye freeze-out is carried out by sampling a thermal (grand-canonical) distribution in the cells on the freeze-out surface.
To start the discussion let us consider the freeze-out procedure in one given cell. Here we will ignore anti-particles which at low collision energies is a good approximation. Given an ensemble of hydrodynamic events, which are generated for example by taking thermal and initial state fluctuations, we can define a probability $P(B)$ to have a given baryon number $B$ in the cell. Then the variance of the baryon number in the cell is given by
\begin{eqnarray}
\sigma^2 &=& \ave{(\delta B)^2 }=\ave{B^2}-\ave{B}^2\non
&=&\sum_B P(B) B^2 - (\sum_B P(B) B )^2
\end{eqnarray}
Now for each run or member of the hydro ensemble we freeze-out by sampling particles according to a Poisson distribution. To keep things general from the start, let us further assume we represent a real baryon with $N_T$ test-particles, so that the baryon number of each test-particle is $Q_B=1/N_T$. Typical freeze-out procedures would correspond to $N_T=1$, i.e. each baryon is represented by one (test)-particle. The mean and the second moment of the baryon number after C-F freeze-out are then given by
\begin{eqnarray}
\acf{B}&=&Q_B \sum_B P(B) \sum_N P_p(N;B/Q) N\non
\acf{B^2}&=&Q_B^2 \sum_B P(B) \sum_N P_p(N;B/Q) N^2
\end{eqnarray}
where, $P_p(N,B/Q) = e^{-B/Q_B} \frac{(B/Q_B)^N}{N!}$ is a Poisson distribution with mean $\ave{N}=B/Q_B$. Here $\acf{\cdots}$ denotes averages after the Cooper-Frye freeze-out. Therefore,
\begin{eqnarray}
\acf{B}&=&Q_B \sum_B P(B) B/Q_B = \ave{B} \non
\acf{B^2}&=&Q_B^2 \sum_B P(B) ((B/Q_B)^2+ B/Q_B) \non 
& = & \ave{B^2} + Q_B \ave{B}
\end{eqnarray}
Obviously, in the second equation $\acf{B^2}\ne \ave{B^2}$ and we obtain an additional contribution to the variance of the baryon number after Cooper-Frye freeze-out, which involves the mean number of baryons in the cell.  
\begin{eqnarray}
\sigma^2_{\mathrm{CF}} = \acf{B^2}-\acf{B}^2=\sigma^2 + Q_B \ave{B}.
\end{eqnarray}
The additional contribution, $Q_B \ave{B}$ scales with $1/N_T$ and is thus suppressed for large numbers of test-particles. 

On the other hand, in case of $N_T=1$, i.e. if each baryon is represented by one test-particle, the extra term may give rise to large extra contribution to the cumulants. For example, if the baryon number of a given fluid cell was already distributed according to a Poisson distribution, we would have
\begin{eqnarray}
\sigma^2_{\mathrm{CF}}=(1+ Q_B)\ave{B}=(1+Q_B)\sigma^2
\end{eqnarray}
so that in case of $N_T=1$ the scaled variance after Cooper-Frye freeze-out would be twice as large than that of the hydro-ensemble.
The above observation can be easily extended to many freeze-out cells and higher cumulants and here we only quote the results for cumulants up to fourth order. The details can be found in the Appendix.
\begin{eqnarray}\label{cumulants}
K_1^{B,CF} &=& \ave{B} = K_1^B \non
K_2^{B,CF} &=& K_2^B + Q_B K_1^B\non
K_3^{B,CF} &=& K_3^B + 3 Q_B K_2^B + Q_B^2 K_1^B \non
K_4^{B,CF} &=& K_4^B + 6 Q_B K_3^B + 7 Q_B^2 K_2^B + Q_B^3 K_1^B
\label{eq:poisson}
\end{eqnarray}
Here $K_n^{B,CF}$ denote the cumulants obtained after Cooper-Frye freeze-out into $N_T = 1/Q_B$ test-particles, and $K_n^B$ denote the true cumulants reflecting the (fluctuating) hydro-ensemble.
Therefore, after Cooper-Frye freeze-out, the the n-th cumulant receives contribution not only from the true n-th order cumulant of the baryon density distribution but from all cumulants of order $m<n$. These extra contributions are suppressed by the number of test-particles, $N_T=1/Q_B$. Therefore, representing a baryon by only one particle at the freeze-out, as it is done commonly when matching various hadronic cascade codes to hydrodynamics, may potentially result in incorrect distribution of the baryon number. We note, however, that in a case where no fluctuations are present in the fluid dynamical fields, i.e. when a constant fluid dynamical background is considered, all fluid dynamical cumulants vanish  $K_i^{B}=0$, and one is left with purely Poissonian cumulants $K_i^{B,CF} = \ave{B}$. This may actually be a desired result expected for random particle number fluctuations. If, on the other hand, one deals with fluctuating fluid dynamics, all or parts of the Poissonian fluctuations are already contained in the hydro ensemble and one is likely to over count the amount of fluctuations by doing a standard C-F freeze-out.\\ 
The above freeze-out based on Poisson sampling does not conserve the total baryon number in a given event. Only in the limit of a large number of test-particles is the baryon number effectively conserved. This situation can be improved by sampling the baryon number (test-particles) from each cell according to a multinomial distribution
\begin{eqnarray}
\lefteqn{P(N_1,\ldots , M_M) = } \non
&& \frac{\left(B_{tot}/Q_{B}\right)!}{N_1! \ldots N_M!}\, p_1^{N_1} \ldots p_M^{N_M}
 \delta_{\sum_{i=1}^M N_i,B_{tot}/Q_{B}}
\label{eq:multinomial}
\end{eqnarray}
Here $B_{tot}$ is the (conserved) baryon number of the entire system, $M$ is the total number of cell of the hyper-surface, $p_{i}=B_{i}/B_{tot}$ with $\sum_{i}^{M} p_{i}=1$, and $B_{i}$ the baryon number in cell $i$ for a given event. 
While this indeed conserves the total baryon number the situation is not much better when one looks at the cumulants of a subsystem (see Appendix \ref{sec:appendix_B} for details)
\begin{eqnarray}
K_1^{B,CF,multi}&=&\ave{B}=K_1^B \non
K_2^{B,CF,multi}&=&K_2^B + Q_B \left( K_1^B - \frac{{K_{1}^{B}}^2+{K_{2}^{B}}}{B_{tot}}\right)
\non
K_3^{B,CF,multi}&=& K_3^B  + 3 Q_B\left( K_2^B -\frac{2 K_1^B K_2^B + K_3^B}{B_{tot}}\right) \non
&& + Q_B^2 \left( K_1^B - 3 \frac{{K_1^B}^2 + K_2^B}{B_{tot}} \right. \non
&& \left.+ 2 \frac{ {K_1^B}^3  + 3 K_1^B K_2^B + K_3^B}{B_{tot}^2} \right)
\label{eq:multi-cumulants}
\end{eqnarray}
Again, we have additional contributions which only disappear in the limit of large number of test-particles, i.e. $Q_{B}\rightarrow 0$.  

\section{Numerical results}

In the following we will present numerical results of a fluid dynamical simulation that entails spatial fluctuations from a first order phase transition and show how these results are changed when a  Cooper-Frye sampling procedure is applied.

\begin{figure}[t]	
\includegraphics[width=0.5\textwidth]{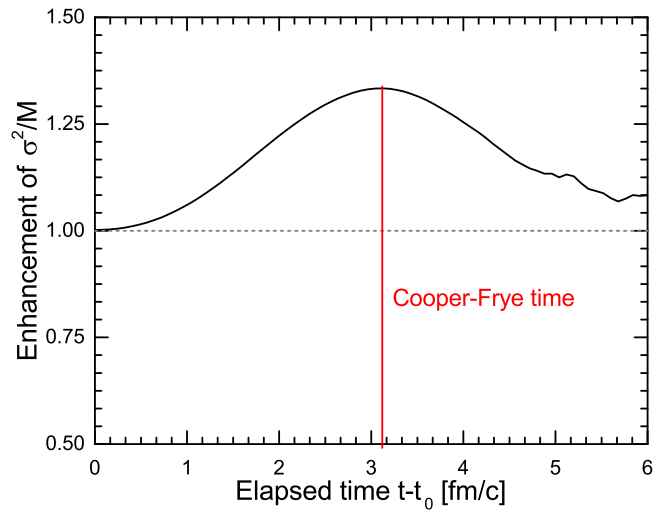}
\caption{[Color online] Time dependence of the enhancement of the scaled variance of the net baryon number, with respect to a simulation without an unstable phase in the EoS. The spatial volume in which the net-baryon number is calculated in each event is fixed by the condition $-0.3 < z < 0.3$ fm, while we integrate over the transverse directions. The enhancement originates from mechanical instabilities due to a first order phase transition. The results shown are for central collisions of Pb nuclei at a fixed target beam energy of $E_{\mathrm{lab}}= 3.5$ A GeV.}\label{f0}
\end{figure}

It has been proposed that the system created in relativistic nuclear collision can undergo spinodal decomposition as it expands rapidly through the spinodal region of the QCD phase diagram \cite{Chomaz:2003dz,Randrup:2003mu}. Recently it was shown that the spinodal decomposition indeed can be observed in realistic fluid dynamical simulations of heavy ion collisions at low beam energies \cite{Steinheimer:2012gc}. The spinodal decomposition occurs as initial seeds of fluctuations are exponentially enhanced as the system expands through the mechanically unstable region  which has been introduced in the equation of state. 
In our present work the seeds of the fluctuations in the fluid dynamical phase are the fluctuations of the initial net baryon number distributions, due to the fact that nuclei are not homogeneous and baryon number stopping therefore varies on an event-by-event basis. These initial fluctuations (calculated with the UrQMD transport model) then are amplified due to the instabilities which occur in the spinodal region of the phase diagram.
The resulting enhancement of the scaled variance of the net baryon number relative to an evolution absent of an unstable region in the equation of state is shown in Fig.~\ref{f0} as function of the total evolution time. This result is for central collisions of lead nuclei and an energy of $E_{\mathrm{lab}}= 3.5$ GeV, where the strongest enhancement is expected  (see \cite{Steinheimer:2012gc}). 

We see from Fig.~\ref{f0} that the enhancement and thus the spinodal clumping is strongest ($\simeq 25\%$) at a time $t=3.0 \, \mathrm{fm/c}$ (see vertical line in figure \ref{f0}). Therefore we will study the effect of the Cooper-Frye particle sampling and  
extract the moments at a point in time. 

In our calculation we have selected events with $N_{{part}}=397$ participating nucleons, to avoid additional contributions from 
$N_{\mathrm{part}}$ fluctuations. Thus the total baryon number in all hydro events will be $B_{tot}=397$.
In our study we will explicitly use a time like hyper-surface $d\sigma_\mu=(dV,0,0,0)$ in order to study fluctuations at a given time. In general all our arguments are also valid for other choices of $d\sigma_\mu$.\\
The sampling procedure used has been explained in \cite{Petersen:2008dd,Huovinen:2012is} and it is constructed to conserve net-baryon number in a single event exactly. This method essentially corresponds to the multinomial sampling discussed in the previous section.

\begin{figure}[t]	
\includegraphics[width=0.5\textwidth]{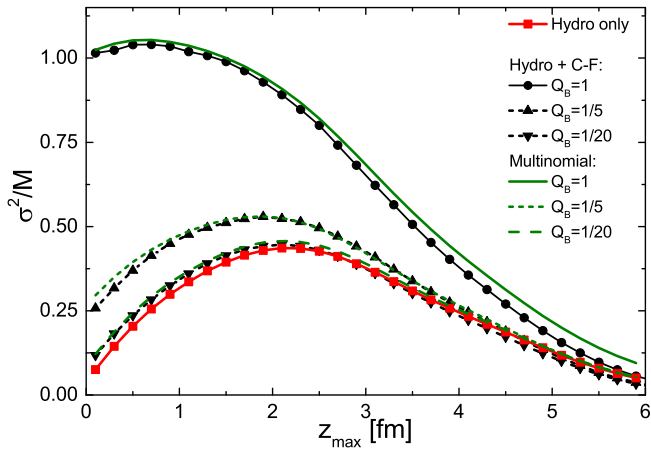}
\caption{[Color online] Variance over the mean of the net-baryon number in a given spatial volume restricted by a maximum z-coordinate value z. The hydro result is obtained at a time t=3 fm/c for collisions of Pb+Pb at a beam energy of $E_{\mathrm{lab}}= 3.5$ A GeV and fixing the number of participants to $N_{\mathrm{part}}=397$.
The red line with square symbols depicts results for the pure hydro simulation before particle production. The black lines with small symbols are the results after particle production via the Cooper-Frye sampling (C-F) with a given number of test-particles per real particle $Q_B=1/N_T$. The green lines are the expected results for a multinomial distribution (eq.(\ref{eq:multi-cumulants})) based on the pure hydro cumulants.}\label{f1}
\end{figure}

Figure \ref{f1} shows the results from the Cooper-Frye sampling of the fluid dynamical simulations at a fixed time and as function of the size of the spatial interval of the sub-system considered. The sub-system is defined as the spatial volume that contains all particles (or hydro matter) with z-coordinates smaller than $|z|<z_{\mathrm{max}}$.

The results of the net baryon number scaled variance, in coordinate space,  for all baryons after the sampling of the the Cooper-Frye equation are shown as black solid line with small circles.
In contrast, the net baryon number variance -- again in coordinate space -- that was directly extracted from the fluid dynamical model is shown in figure \ref{f1} as red line with squares.

\begin{figure}[t]	
\includegraphics[width=0.5\textwidth]{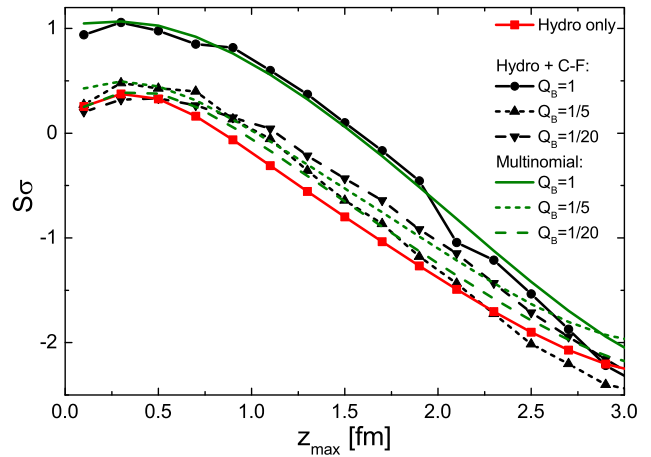}
\caption{[Color online] Same as in figure \ref{f1} but for the Skewness.}
\label{f4}
\end{figure}

As expected one obtains a significantly different scaled variance due to the contribution from the freeze-out procedure, as shown in Eq.~\eqref{eq:multi-cumulants}. Indeed, if we use equation (\ref{eq:multi-cumulants}) to add the contribution $\sim Q_{B}$ to the variance and set $Q_{B}=1$, we obtain the solid green line which agrees very well with the result from the numerical sampling. The same is true also for the Skewness, shown in Fig.~\ref{f4}.

If we carry out the C-F freeze-out using test-particles, i.e. $Q_{B}<1$, the resulting scaled variance and skewness quickly approach the true hydro result. The results of the C-F sampling using an increasing number of test-particles is shown as black dashed lines with symbols in Figs.~\ref{f1} and \ref{f4}. The corresponding results using Eq.~\eqref{eq:multi-cumulants} for $Q_{B}<1$ are show as the green dashed lines. Again for the scaled variance we find a very good agreement of the C-F test particle sampling and the multinomial cumulants, Eq.~\eqref{eq:multi-cumulants}. 
For the skewness we have plotted only results up to $z=3$ since for higher values of $z$ already the pure hydro results (red line) is not stable within the statistics of our calculations.

\section{Discussion}

We have shown that the finite particle sampling leads to additional contributions to the local fluctuations of particle numbers, contributions which are not present in the fluid dynamical simulation. Using test-particles these additional contribution can be suppressed. 
However, there are various subtleties which one needs to take into account:
\begin{enumerate}
\item
We would like to stress that we do not claim that the extra contributions arising from C-F sampling are a priory wrong in every situation. For example, in case of a standard (non-fluctuating) fluid dynamical simulation it is probably reasonable to assume that the particle number fluctuates locally according to a Poissonian (or multinomial for a globally conserved charge) distribution. Therefore the Cooper-Frye sampling would correctly reproduce the local thermal fluctuations. 
This is also true if the 'clumps' of baryonic matter, produced by e.g. the spinodal decomposition, are of macroscopic size i.e. contain a very large number of particles. In this case the cumulants are then dominated by the fluctuations of these clumps, as one can easily see from Eq.\eqref{eq:poisson}\footnote{Here one should keep in mind that for an increasing clump size, i.e. in the presence of a strong long range correlation, the first order cumulant only increases as $\ave{B}$ while the second order cumulant will increase as $\ave{B^2}$} and the extra contribution from the C-F sampling are subleading. Or in other words, as long as the variations of the baryon number in the fluid are "long range" the addition of local, independent fluctuations from the freeze-out is most likely reasonable and in any case a subleading contribution.
\item 
The situation gets more tricky if one deals with initial fluctuations and their possible enhancement due to spinodal instabilities in nuclear collisions. Here the clumps contain only a few baryons and thus do not dominate the cumulants and the contribution from the C-F sampling are non-negligible. Therefore, it is not clear if one can treat the clumps of baryonic matter as sufficiently large to consider them macroscopic fluctuations. If this is not the case then the application of standard fluid dynamics is questionable. A possible solution of this would be the application of fluctuating fluid dynamics. 
\item
In a model where the thermal fluctuations are already an integrated part of the fluid dynamics evolution, the additional fluctuations generated by the C-F freeze-out prescription would be unphysical and need to be removed. However, it is not clear that such a numerical description will give stable result if the particle number in a given cell is so small that the local thermal fluctuations are of the same order or even bigger than the mean value.
\item 
We have proposed to use test-particles in order to suppress the additional fluctuations from C-F if needed. This can be easily implemented using existing codes. One could of course argue that one follows a (micro)-canonical freeze-out prescription, where the baryon number in each cell is exactly conserved. This, however, will likely also require the use of test-particles for the hadronic transport, since the baryon number is some cells is likely to be smaller than unity. And even if one combines several cells, it is not clear that one always has an integer baryon number for a region where the temperature is sufficiently constant. Again, using test-particles is a simple way to  accommodate the non-integer baryon number of a freeze-out region in the hadronic transport.
\item We note that the use of test-particles in hadronic transport will suppress correlation from resonance decays, which, if essential, will have to be implemented explicitly by propagating the two particle distribution function. This will complicate the numerical implementation  of the transport evolution considerably. 
Also, it is essential that the test-particles carry a fractional baryon-number $Q_{B}$. Therefore, so called parallel ensemble methods used for a more efficient transport code are not possible.  
\item Here we have only considered the fluctuations of the baryon number. However, it is rather clear from the discussion, that the C-F method will affect the fluctuations of other conserved quantities such as energy and momentum in an analogous fashion. To which extend this affects the outcome of flow fluctuation observables would be interesting to study. 
\item As we have pointed out fluid dynamics is based on the propagation of energy momentum and conserved currents. Usually one only considers the time evolution of the net baryon number current. By the construction of the equation of state one also implies certain values for the isospin and net strangeness, usually they are assumed to be exactly zero. 
In such a case, for a given local net baryon density, the net proton and net neutron numbers are exactly determined by the EoS in the fluid dynamical picture. In reality of course not only the net baryon number may fluctuate but the number of any particle species within a given cell may fluctuate independently. This is naturally included in the finite number sampling. In the fluid dynamic model one would have to explicitly include dynamically the thermal fluctuations of all particle species which is impractical. As a result the fluid dynamical model will always obtain proton number fluctuations which are exactly correlated with the baryon number fluctuations. E.g. for vanishing baryon fluctuation $\sigma_B^2=0$ one still obtains $\sigma_p^2 = 0$ instead of $\sigma_p^2= 1/2 \left\langle p \right\rangle$ \cite{Begun:2005ah}  if one allows the iso-spin fluctuate while fixing the baryon number.
\item 
In this paper we have ignored the anti-baryons, which is a reasonable approximation for collision energies $\sqrt{s}\leq 20 \, \rm{GeV}$, where the observed anti-proton to proton ratio is very small. It would be interesting to extend the presented test-particle freeze-out to include also anti-particles. The challenge in this case would be to conserve the baryon number, i.e. the difference between baryons and anti-baryons, in a given cell while still allowing for the sum to fluctuate. 
\end{enumerate}

\section{Summary}

We have shown that applying the commonly used finite particle sampling method for the Cooper-Frye particle production will introduce an additional contribution to the particle number distributions. Only in the very specific case of the production of (locally) uncorrelated particles will this method give the physically correct moments of the particle number distributions.
In the case of a fluctuating fluid dynamics model, the finite particle sampling will introduce unphysical contributions, which can be suppressed by applying  test particle sampling.
We have shown how using test particle sampling one can recover the original susceptibilities from the fluid dynamical simulation. We found that for the specific, low beam energies under investigation, 20 test particles per real particles will give a result reasonably close to the original second and third order susceptibilities.
We have derived formulas, based on a multinomial distribution, which, in the absence of anti-baryons, reproduces the cumulants obtained from a numerical C-F sample including global baryon number conservation very well. Actually it might be worthwhile to explore if these formulas may be used to test and validate the various freeze-out schemes which address global particle number conservation.\\  
Finally we have presented results from a realistic simulation of the spinodal instabilities due to a first order phase transition. These results indicate that the effect of the spinodal clumping on the fluctuations is in fact much smaller than what one would expect from thermal (Poissonian) fluctuations, at least regarding the scaled variance and skewness. This comes as a result of the small system size of the nuclear collisions and the fact that the created clumps of baryonic matter cannot be considered macroscopic objects, thus the cumulants are dominated by the local thermal/random fluctuations of the baryon number.

\section{Acknowledgments}
We would like to thank the Institute for Nuclear Theory where 
this paper was initiated during the program ``Exploring the QCD Phase Diagram through Energy Scans''. In addition V.K. would like thank U. Heinz, S. Jeon, B. Schenke, and A. Wergiluk for discussions. V.K. was supported by the Office of Nuclear Physics in the US Department of Energy's Office of Science under Contract No.
DE-AC02-05CH11231. J.S. was supported by the Helmholtz Center for Heavy Ion Research GSI and HIC for FAIR. This work also received support within the framework of the Beam Energy Scan Theory (BEST) Topical Collaboration. 

\appendix
\section{Poisson sampling}
\label{sec::app_a}
Here we calculate the resulting cumulants after Cooper-Frye freeze-out using $N_T$ test-particles.  Again we denote the baryon number per test-particle by $Q_B=1/N_T$ so that the cumulants of the baryon number, $K_n$, are related with those for the test particles, $k_n$ by
\begin{eqnarray}
K^{B}_n = Q_B^n k_n
\end{eqnarray}

Cumulants are best obtained from the cumulant generation function $g(z)$. Given the probability $P(N)$ to have $N$ test-particles, 
\begin{equation}
g(z) = \ln\left(\sum_N P(N) e^{z N} \right).
\end{equation}
The cumulant of the test particle distribution of order $n$, $k_n$ is obtained from the generating function by
\begin{equation}
k_n = \frac{d^n}{dz^n} \left. g(z) \right|_{z=0}
\end{equation}
Given $n_c$ cells in which particles are frozen out and the probability $P(B_i,\cdots,B_{n_c})$ to have a baryon number of $B_i$ in cell $i$, the probability $P(N)$ to have N test particles is given by
\begin{eqnarray}
P(N) & = & \sum_{\{B_1,\cdots,B_{n_c}\}} P(B_i,\cdots,B_{n_c}) \nonumber \\ 
 &\times& \sum_{k_1} \exp(B_1/Q_B) \frac{(B_1/Q_B)^{k_1}}{k_1!} \cdots  \nonumber \\
 &\cdots &\sum_{k_{n_c}} \exp(B_{n_c}/Q_B) \frac{(B_{n_c}/Q_B)^{k_{n_c}}}{k_{n_c}!} \nonumber \\
&&\delta_{k_1+\cdots+k_{n_c},N}
\end{eqnarray}
Here, $\{B_1,\cdots,B_{n_c}\}$ represents the set of baryon numbers in cell $1$ to $n_c$, and $P(B_i,\cdots,B_{n_c})$ the probability that such a configuration is present in the ensemble of hydrodynamic runs. Furthermore $\sum_i B_i = B$ is the baryon number of a given configuration.
The generating function $g(z)$ is then easily determined
\begin{eqnarray}
g(z)&=&\ln \left[\sum_{\{B_1,\cdots,B_{n_c}\}}P(B_i,\cdots,B_{n_c}) \times \right. \non 
&& \left. M(z;B_1/Q_B)\cdots M(z,B_{n_c}/Q_B) \right],
\label{eq:generate_2}
\end{eqnarray}
where 
\begin{eqnarray}
\lefteqn{M(z,B_i/Q_B) =} \non 
&& \sum_k \exp(B_{i}/Q_B) \frac{(B_{i}/Q_B)^{k}}{k!} \exp(z k) \non 
&=& e^{ B_i/Q_B (e^t - 1)} = e^{ B_i y(z)} 
\end{eqnarray}
is the moment generating function for a Poisson distribution with mean $\nu = B_i/Q_B$. Here, $y(z)=\frac{e^z-1}{Q_B}$ so that $y(0)=0$. Inserting the explicit form of the moment 
generating functions into Eq.~\ref{eq:generate_2} we get
\begin{eqnarray}
g(z) &=&\ln\left( \sum_{\{B_1,\cdots,B_{n_c}\}}P(B_i,\cdots,B_{n_c})e^{\sum_i B_i y(z)} \right)\non
&=& \ln \sum_B P(B) e^{B y(z)} \non
&=& G_B\left(y(z)\right)
\label{eq:generate_final}
\end{eqnarray}
where  $G(x)$ is the cumulant generating function of the  the baryon number distribution $P(B)$.
Noting that $\left. \frac{d^n}{d z^n} y(z) \right|_{z=0} = 1/Q_B$ for $n\geq 1$,   
the baryon cumulants after Cooper-Frye freeze-out are given by
\begin{eqnarray}
K_1^{B,CF} &=& Q_B k_1 =  Q_B \left.\frac{d}{d z} g(z)\right|_{z=0}\non 
&=& \left. G'(y(z))y'(z)\right|_{z=0}=\ave{B}= K_1^B \non
K_2^{B,CF} &=& Q_B^2 k_2 = Q_B^2\frac{d^2}{d z^2} g(z)\non 
&=& Q_B^2 \left[ G''(0) \left(y'(0)\right)^2 + G'(0) y''(0) \right] \non
&=& \ave{(\delta B)^2} + Q_B \ave{B} = K_2^B + Q_B K_1^B\non
K_3^{B,CF} &=& K_3^B + 3 Q_B K_2^B + Q_B^2 K_1^B \non
K_4^{B,CF} &=& K_4^B + 6 Q_B K_3^B + 7 Q_B^2 K_2^B + Q_B^3 K_1^B
\label{eq:appndx_all_cumulants}
\end{eqnarray}
Here $C_n^{B,CF}$ denote the cumulants obtained after Cooper-Frye freeze-out into $N_T = 1/Q_B$ test-particles, and $K_n^B$ denote the true cumulants reflecting the (fluctuating) hydro-ensemble.

\section{Multinomial sampling}
\label{sec:appendix_B}
Another way to handle the Cooper-Frye freeze-out would be to sample the  baryon number with a multinomial distribution. In the limit that the baryon number in the individual cells is small compared to the total baryon number, this procedure is very similar to the Poisson sampling discussed in the previous section, with the added advantage that the total baryon number is conserved. Let us assume that the entire freeze-out hyper-surface is comprised of of $M$ cells, where on the hydro side each cell $i$ has a baryon number of $B_i$. The total baryon number $B_{tot}$ is then given by  $\sum_{i=1}^M B_i = B_{tot}$. The probability distribution from which to sample the number of test-particles $N_i$ in the cells $i$ is then given by the multinomial distribution
\begin{eqnarray}
P(N_1,\ldots , M_M) &=& \frac{\left(B_{tot}/Q_{B}\right)!}{N_1! \ldots N_M!} p_1^{N_1} \ldots p_M^{N_M}
\non 
&& \delta_{\sum_{i=1}^M N_i,B_{tot}/Q_{B}}
\end{eqnarray}
with $p_i = \frac{B_i}{B_{tot}}$. Clearly this prescription ensure that the {\em total} baryon number after sampling is the same as that prior to freeze-out. The algorithms described in \cite{Petersen:2008dd,Huovinen:2012is} which try to ensure global baryon number conservation baryon number are very similar to this multinomial sampling. While the sampling a multinomial distribution ensures the global conservation of the baryon number, it still fails to faithfully map the local fluctuations of the baryon number from hydrodynamics. To see this let us consider as subset of the total freeze-out hyper-surface, given by a number of cells $m$ with $m<M$. Let us further denote the baryon number on the hydro side of this subset by $B=\sum_{i=1}^m B_i$. Since we are only interested in the distribution of the baryon number in the subset of cell the above multinomial distribution reduces to a binomial distribution. Thus the baryon number of the subset after freeze-out $b = \sum_{i=1}^m b_i$ is governed by
\begin{eqnarray}
P(b) = \frac{B_{tot}!}{b! (B_{tot}-b)!} p_m^{b}(1-p_m)^{B_{tot}-b}
\end{eqnarray}
where the binomial probability $p_m$ is given by the sum of the probabilities of the cells of interest, $p_m=\sum_{i=1}^m p_i$.
The corresponding distribution for the number of test-particles, $P(N)$, in the subset is then given by
\begin{eqnarray}
P(N) &=& \frac{\left(B_{tot}/Q_B\right) !}{N! \left(B_{tot}/Q_B - N\right)!}
p_m^{M}(1-p_m)^{B_{tot}/Q_B-N} \non
\end{eqnarray}

Given the above distribution the various cumulants of the baryon number distribution in the subsystem are
\begin{eqnarray}
K_1^{B,CF,multi}&=&\ave{B}=K_1^B \non
K_2^{B,CF,multi}&=&K_2^B + Q_B \left( K_1^B - \frac{{K_1^B}^2+ K_2}{B_{tot}}\right)
\non
K_3^{B,CF,multi}&=& K_3^B  + 3 Q_B\left( K_2^B -\frac{2 K_1^B K_2^B + K_3^B}{B_{tot}}\right) \non
&& + Q_B^2 \left( K_1^B - 3 \frac{{K_1^B}^2 + K_2^B}{B_{tot}} \right. \non
&& \left.+ 2 \frac{ {K_1^B}^3  + 3 K_1^B K_2^B + K_3^B}{B_{tot}^2} \right).
\non
\end{eqnarray}

Again,  $\ave{B}=K_{1}^{B}$ is the mean baryon number and $K_2^B$ and $K_3^B$ are the cumulants based on the ensemble of the hydro events for the subsystem under consideration. $B_{tot}$ is the total baryon number of the entire system which is the same in each event in the ensemble. We note, that if we consider the entire event, then $K_2^B=K_3^B=0$ and $\ave{B}=B_{tot}$ and, consequently, $K_2^{B,CF,multi}=K_3^{B,CF,multi}=0$. So indeed, global baryon number conservation is achieved in case of the multinomial freeze-out model.
Furthermore, in the limit  where the subset is small compared to the entire system, $K_i^B \ll B_{tot}$ we recover the result based on the Poisson distribution, Eq.~\ref{eq:appndx_all_cumulants}.


\end{document}